\documentclass[conference, 10pt]{IEEEtran}
\IEEEoverridecommandlockouts
\usepackage{cite}
\usepackage{amsmath,amssymb,amsfonts}
\usepackage{array}
\usepackage[ruled,linesnumbered,lined]{algorithm2e}
\usepackage{graphicx}
\usepackage{textcomp}
\usepackage{xcolor}
\usepackage{bm}
\usepackage{titlesec}
\def\BibTeX{{\rm B\kern-.05em{\sc i\kern-.025em b}\kern-.08em    T\kern-.1667em\lower.7ex\hbox{E}\kern-.125emX}}

\begin{document}
	\title{Individual Channel Estimation for  Beyond Diagonal Reconfigurable Intelligent Surfaces}
	\author{
		\IEEEauthorblockN{Kangchun Zhao and Yijie Mao}
		\IEEEauthorblockA{
		School of Information Science and Technology, ShanghaiTech University, Shanghai 201210, China \\		
			Email: \{zhaokch12022, maoyj\}@shanghaitech.edu.cn}
		\\[-1.5 ex]
		 
		\\[-3.5 ex]		
	}
\maketitle
\thispagestyle{empty}
\pagestyle{empty}
\begin{abstract}
Beyond Diagonal Reconfigurable Intelligent Surfaces (BD-RIS) has emerged as a promising evolution of RIS technology.
By enabling interconnections between RIS elements, BD-RIS architectures offer greater flexibility in wave manipulation compared to traditional diagonal RIS designs. 
However, these interconnections introduce new research challenges for channel estimation, making existing approaches developed for conventional diagonal RISs ineffective and significantly increasing pilot overhead.
To address these challenges, we propose a novel individual channel estimation framework that separately estimates the BS-RIS channel, which typically remains static over time, and the RIS-user channels, which vary rapidly due to user mobility. 
Specifically, we develop a full-duplex (FD) approach to estimate the BS-RIS channel by leveraging its inherent sparsity.
Following this, the RIS-user channels are estimated using a least squares (LS) approach. 
Numerical results demonstrate that the proposed framework achieves significantly higher channel estimation accuracy, particularly when the number of RIS elements is large, while substantially reducing pilot overhead compared to conventional cascaded channel estimation methods.
\end{abstract}


\section{Introduction}
Beyond diagonal reconfigurable intelligent surfaces (BD-RIS) have emerged as an evolution of conventional diagonal-RIS (D-RIS) technology\cite{di2020smart}. 
Unlike conventional D-RIS, where each element is adjusted independently through a diagonal phase-shift matrix, BD-RIS introduces interconnections among RIS elements\cite{li2023reconfigurable}, allowing the resulting phase-shift matrix (also known as the scattering matrix) to be a more general, non-diagonal structure.
For example, for a fully connected RIS architecture, the scattering matrix becomes a full matrix.
By enabling inter-element connections, BD-RIS significantly enhances the ability to manipulate wireless channels, leading to improvements in channel gain, spectral efficiency, and communication coverage\cite{shen2021modeling}, etc.

Channel estimation has become a critical challenge in BD-RIS-aided communication networks. 
Inaccurate channel state information (CSI) at the base station (BS) leads to inaccurate downlink passive beamforming, resulting in degraded system performance and increased inter-user interference. 
Due to the passive nature of BD-RIS, it cannot actively transmit or decode information signals, making the estimation of both the BS-RIS channel and the RIS-user channels particularly difficult. 
Furthermore, the inter-connections among RIS elements in BD-RIS significantly increase the dimension of the cascaded BS-RIS-user channel compared to conventional D-RIS, thereby introducing substantially higher pilot overhead.

The study of channel estimation for BD-RIS is still in its early stages. 
To date, only a few works \cite{li2024channel,de2024channel,sokal2024decoupled,wang2025low} have investigated this issue, and all are limited to the estimation of the cascaded BS-RIS-user channel.
Specifically, in \cite{li2024channel}, a least square (LS)-based method is proposed, which leverages DFT or Hadamard matrices to design the scattering matrix.
However, the pilot overhead of this LS-based method is proportional to the square of the number of BD-RIS elements for fully connected RIS, which is rather high.
Based on\cite{li2024channel}, in \cite{de2024channel}, two channel estimation methods are proposed based on tensor decomposition.
However, compare with the method in \cite{li2024channel}, though it reduces the pilot overhead, the estimation error is sacrificed.
In \cite{sokal2024decoupled}, a decoupled channel estimation method is proposed to separately estimate different parts of the cascaded channel based on the LS
results.
However, it requires to estimate the cascaded channel first, so the pilot overhead is still heavy.
In \cite{wang2025low}, the author proves that the cascaded channel for BD-RIS can be recovered by estimating the   cascaded
channel associated with one pair of BS antenna and BD-RIS element, which reduces the pilot overhead to the same order as D-RIS.
However, such approach still requires frequent estimation of the cascaded channel. 
To the best of our knowledge, all existing works on BD-RIS channel estimation rely on cascaded channel estimation. 
There is currently no method that directly estimates the BS-RIS channel and the RIS-user channels separately.

To bridge this research gap, in this paper, we propose an individual channel estimation method to estimate the BS-RIS channel and RIS-user channel, respectively.
In the estimation of BS-RIS channel, by exploiting the sparsity of the channel, we propose a full-duplex (FD) method to estimate the elevation angles at the BS, the elevation and azimuth angles at the BD-RIS, and the channel gains, separately. 
Next, given the estimated BS-RIS channel, we use a LS-based algorithm to estimate the rest RIS-user channels.
Numerical results demonstrate that the proposed framework significantly improves channel estimation accuracy, especially as the number of RIS elements increases, while also substantially reducing pilot overhead compared to conventional cascaded channel estimation methods.

\section{System Model}
\label{sec:: sys}
Consider a  narrowband time division duplex (TDD) BD-RIS-assisted multi-user multiple-input single-output (MU-MISO) network consisting of a base station (BS) with $M$ antennas, a fully-connected BD-RIS with $N$ elements, and $K$ single-antenna users. 
\begin{figure}[tb]
\centerline{\includegraphics[width=0.38\textwidth]{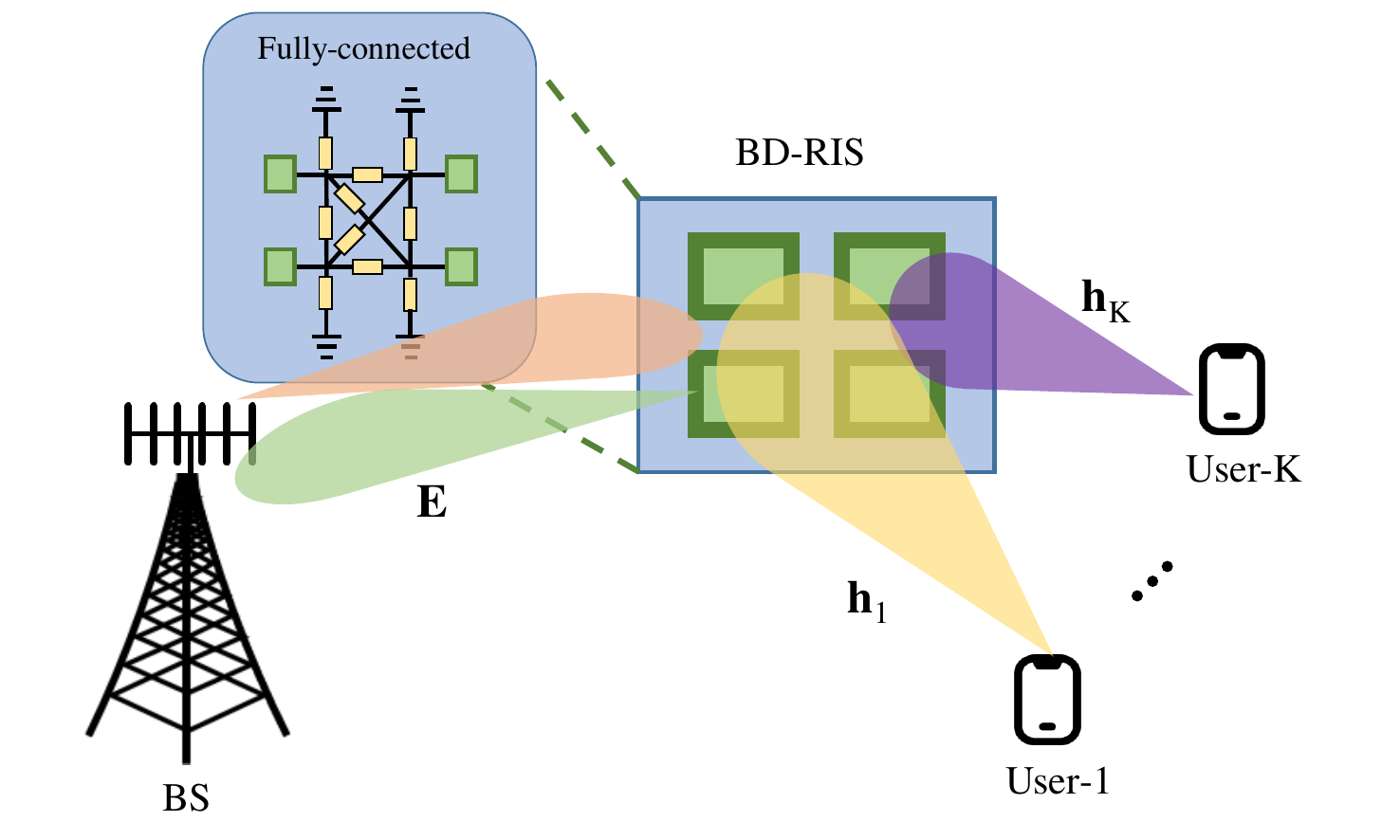}}
\caption{A BD-RIS aided multi-user communication network.}
\label{fig1}
\end{figure}
As the direct channel between the BS and  user-$k$ can be easily estimated using conventional approaches\cite{swindlehurst2022channel}, in this work, we focus on estimating the cascaded BS-RIS-user channel and ignore the BS-user links.
The channels between the BS and the BD-RIS, the BD-RIS and user-$k$ are respectively denoted by $\mathbf{E}\in\mathbb{C}^{M\times N}$,  and $\mathbf{h}_k\in\mathbb{C}^{N\times 1}$.
Suppose that a uniform linear array is equipped at the BS and a uniform planar array (UPA) is equipped at the BD-RIS. By adopting the widely considered narrowband Saleh-Valenzuela (S-V) channel model\cite{hu2018super}, the BS-RIS and the RIS-user channels are respectively denoted as:
\vspace{-1.5mm}
    \begin{equation}
    \small
    \begin{split}       \mathbf{E}&=\sum^{L}_{l=1}\alpha_{l}\mathbf{b}\left( \iota_{b,l}\right) \mathbf{a} \left( \iota_{r,l},\phi_{r,l} \right)^T,\label{channel::BS-RIS}\\ 
    \mathbf h_k &= \sum^{U_k}_{l=1}\alpha_{k,l}\mathbf{a} \left( \iota_{k,l},\phi_{k,l} \right) , \forall k\in\mathcal{K}.
    \end{split}
\end{equation}
where $L(U_k)$ is the number of channel paths between the BS and the BD-RIS (the BD-RIS and  user-$k$), $\alpha_{l}(\alpha_{k,l})$ denotes the complex channel gain of the $l$-th path, $\iota_{b,l}$ denotes  the elevation angle at the BS, $\iota_{r,l}\left(\iota_{k,l} \right)$ and $\phi_{r,l}\left(\phi_{k,l} \right)$ denote  the elevation and azimuth angles  associated with the BS-RIS channel and RIS-user channel, respectively.
$\mathbf{b}\left( \iota\right)\in\mathbb{C}^{M\times1}$ denotes the  steering vector at the BS, which is typically modeled as:
    \begin{equation}
    \small
    \mathbf{b}\left( \iota\right)=e^{-j2\pi d\cos(\iota)\mathbf{m}/\lambda},\label{vector::BS}
\end{equation}
where  $\mathbf{m}=\left[0,1,\ldots,M-1\right]^T$, $\lambda$ is the carrier wavelength and $d$ is the distance between two adjacent antennas, $d=\frac{\lambda}{2}$.
$\mathbf{a}\left( \iota,\phi\right)\in\mathbb{C}^{N\times1}$ denotes the  steering vector at the BD-RIS.
For a $N~(N=N_1\times N_2)$-element UPA, $\mathbf{a}\left( \iota,\phi\right)$ is given as 
 \begin{equation}
 \small
    \mathbf{a}\left( \iota,\phi\right)=e^{-j2\pi d\sin(\phi)\cos(\iota)\mathbf{n}_1/\lambda}\otimes e^{-j2\pi d\sin(\iota)\mathbf{n}_2/\lambda},\label{vector::RIS}
\end{equation}   
where  $\mathbf{n}_1=\left[0,1,\ldots,N_1-1\right]^T$ and $\mathbf{n}_2=\left[0,1,\ldots,N_2-1\right]^T$.
Our goal is to estimate  $\mathbf{E}$ and $\mathbf{h}_k$,  $ \forall k\in\{1, \ldots, K\}$. 
The BD-RIS is fully connected, whose reflection matrix  at $t$-th time slot is $\bm{\Theta}_t$, satisfying $\bm{\Theta}_t^H\bm{\Theta}_t = \mathbf I_{N}$.

\textit{Conventional approach--cascaded channel estimation}:
Existing works\cite{li2024channel,de2024channel} directly estimate the cascaded BS-RIS-user channel by sending pilots from the users to the BS. Specifically, all users simultaneouly send pilot signals to the BD-RIS, and then to the BS. Let $x_{k,t}$ denote the pilot sent from user-$k$, the received signal at the BS from all users at the $t$th time slot are given as: 
\vspace{-2.5mm}
\begin{align}
\mathbf{y}_{t}&=\sum^{K}_{k=1}\mathbf{E}\bm{\Theta}_t\mathbf{h}_kx_{k,t} +\mathbf{n}_t,\label{signal::uplink}\\
&=\sum^{K}_{k=1}\underset{\mathbf{H}_k}{\left(\underbrace{\mathbf h^T_k\otimes\mathbf E}\right)}\text{vec}\left(\bm{\Theta}_t\right)x_{k,t}+\mathbf{n}_t,\label{signal::uplink3}
\end{align}
where the operator  $\text{vec}(\cdot)$ makes a matrix vectorization thereby  $\text{vec}\left(\bm{\Theta}_t\right)\in\mathbb C^{N^2\times 1}$,   $\mathbf{n}_t\in\mathbb{C}^{M\times 1}$ is the additive noise and $\mathbf{H}_k\in\mathbb C^{M\times N^2}$ is the cascaded channel to be estimated. 
To enable the BS to estimate $\mathbf H_k$, each user must transmit a sufficient number of pilots during the training phase.
As pointed out in \cite{li2024channel}, at least $N^2$ pilots must be received at the BS to successfully estimate $\mathbf H_k$ for a fully connected BD-RIS-assisted cascaded link. 
Since only one pilot is transmitted per time slot, a total of $T \geq N^2$  time slots are required to complete the pilot transmission. 
It is clear that when $N$ is large, the pilot overhead of this approach becomes extremely high. 
This pilot overhead burden arises from the interconnections among elements in BD-RIS.
In contrast, with a conventional D-RIS, the cascaded BS–RIS–user channel simplifies to estimate $\mathbf H^{D}_k=\mathbf E\text{diag}\left(\mathbf h_k\right)\in\mathbb C^{M\times N}$,  only $T\geq N$ pilots are needed to estimate  $\mathbf H^{D}_k$ at the BS.
Hence, cascaded channel estimation for BD-RIS in rather challenging due to much higher pilot overhead than D-RIS.
\vspace{-3mm}
\section{Individual Channel Estimation Framework}
\begin{figure}[tb]
\centerline{\includegraphics[width=0.48\textwidth]{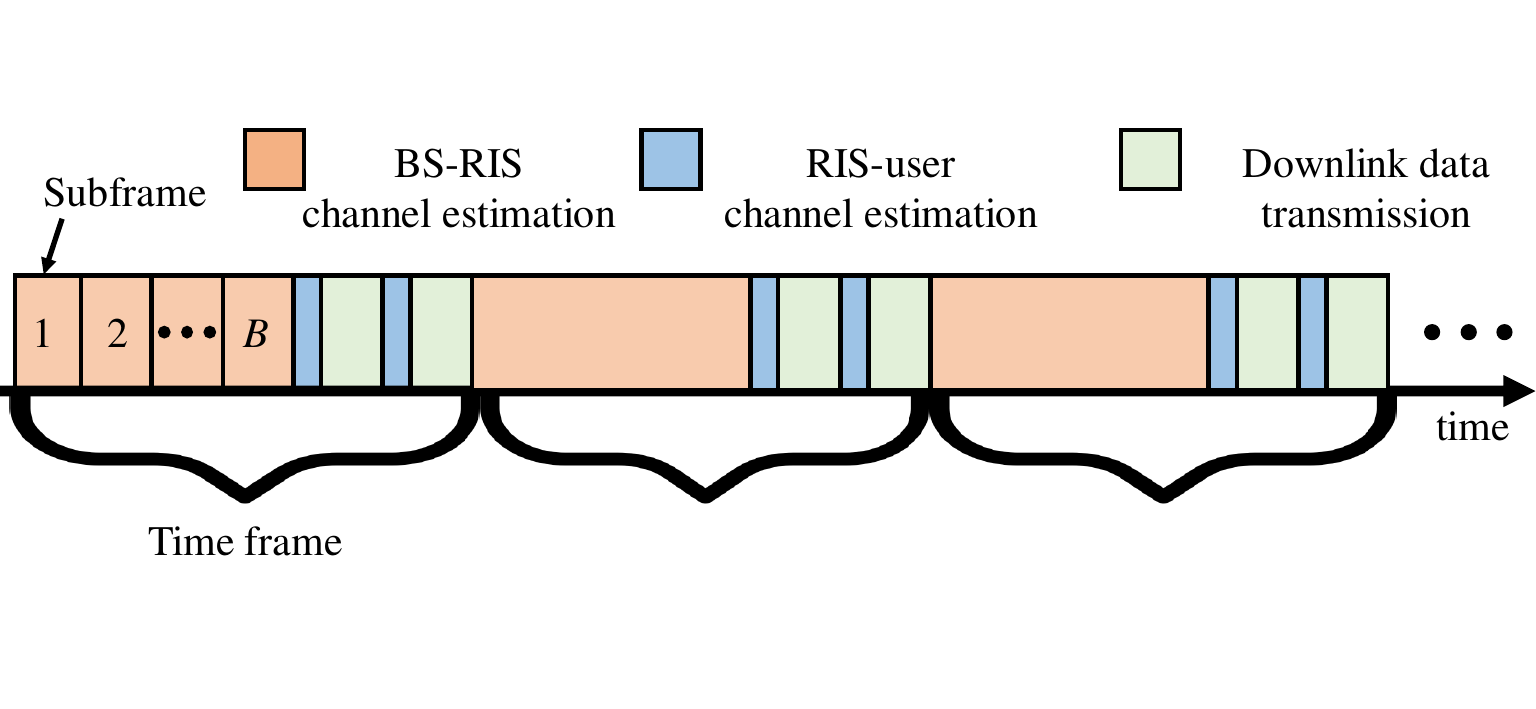}}
\vspace{-3mm}
\caption{BD-RIS channel estimation structure.}
\label{fig2}
\vspace{-3mm}
\end{figure}
To address the pilot overhead introduced by cascaded channel estimation, we propose a novel individual channel estimation framework for BD-RIS in this section.
Instead of directly estimating $\mathbf H_k$, the proposed scheme separately estimates the BS-RIS channel $\mathbf E$ and the RIS-user channels $\mathbf h_k$, $\forall k$. 
The key motivation behind this approach is that the BS-RIS channel $\mathbf E$ remains relatively stable, as the positions of the BS and the BD-RIS are typically fixed. 
Thus, $\mathbf E$ only needs to be estimated infrequently over extended periods. 
In contrast, the RIS-user channels $\mathbf h_k$, $\forall k$ vary more rapidly due to user mobility, requiring more frequent estimation. 

 The proposed individual channel estimation framework is illustrated in Fig. \ref{fig2}.
 The time scale is divided hierarchically into time frames, subframes, and time slots. 
 In one time frame, BS-RIS channel $\mathbf E$ is estimated once, which requires $B$ subframes. 
 The RIS-user channel is estimated multiple times, each time requires $C$ subframes.  
 Our channel estimation procedure involves two major steps within each time frame:

\begin{itemize} 

\item Step 1: At the first $B$ subframes of each time frame, the BS-RIS channel $\mathbf E$ is estimated.  
Due to the passive reflection nature of BD-RIS, it cannot transmit or decode wireless signals on its own, making individual channel estimation challenging. 
To address this limitation, inspired by the FD estimation schemes proposed in \cite{hu2021two} for conventional D-RIS, we adopt a similar FD-enabled BS for estimating the BS-RIS channel $\mathbf E$. 
Specifically, the BS sends pilots toward the BD-RIS and simultaneously receives the reflected signals. 
The BS then estimates $\mathbf E$ based on the received signals.
$\mathbf E$ is assumed to remain constant throughout the entire time frame. 

\item Step 2: Within $C$ subframes, the RIS-user channels $\mathbf h_k$, $\forall k$  are estimated. 
Users send pilots to the BD-RIS, which reflects the signals to the BS for channel estimation. 
$\mathbf h_k$ is assumed to remain constant within $C$ time subframes. 
\end{itemize}
The details of these two steps are provided in the following subsections.

\subsection{BS-RIS Channel Estimation}
 In this step, the BS operates in the FD mode with the aim of estimating the BS-RIS channel $\mathbf E$. 
According to (\ref{channel::BS-RIS}), it is obvious that once we obtain the complex channel gain $\alpha_l$, the elevation angle at the BS $\iota_{b,l}$, and the azimuth (elevation) angle at the BD-RIS $\iota_{r,l}(\phi_{r,l})$, for all channel paths between the BS and BD-RIS, we could recover $\mathbf E$.
Here, we assume a bistatic BS configuration, where different antennas are used for transmission and reception. 
Specifically, the $M$ antennas at the BS are partitioned into two groups: $M_T$ antennas are dedicated to transmitting pilots toward the BD-RIS, while the remaining $M_R = M - M_T$ antennas are used to receive the reflected signals.
The BS then estimates $\mathbf E$ based on the received pilot signals. 
It is worth noting that the bistatic setup, with separate transmit and receive antennas, provides spatial isolation, which effectively reduces radio self-interference\cite{zhou2024individual}.

As illustrated in Fig. \ref{fig2}, in each time frame,  the BS-RIS channel $\mathbf E$ is estimated only once, requiring $B$ subframes, and therby $BT$ time slots in total to complete the estimation process.
According to \cite{emmanuel2006robust}, to
 estimate a $l$-sparse complex signal with $n$ dimension we require a  pilot overhead on the scale of $\mathcal{O}\left(l\log(n)\right)$.
To estimate the elevation angles at the BS using $BT$ slots, only if $T\geq L$, $B \geq \log(M)$, can we successfully  estimate $\iota_{b,l}$ at the BS.
To estimate the azimuth and elevation angles at the BD-RIS with pilot overhead of $BT$,  $B\geq \log(N^2)$.
In each subframe, the BS transmits the pilot matrix $\mathbf S$ satisfying $\mathbf S^H\mathbf S=\mathbf I$ to the BD-RIS through the channel $\mathbf{E}_T\in\mathbb C^{M_T\times1}$, the BD-RIS then reflects the signal back to the BS via the $\mathbf{E}_R\in\mathbb C^{M_R\times1}$.
During the $b$-th subframe, the signal received at the BS via the BS-RIS and RIS-BS links is given as: 
\begin{align} 
\mathbf{Y}_b=\sqrt{P}\mathbf{E}_R\bm{\Theta}_{b}\mathbf{E}_T^T\mathbf S+\mathbf N_b,\label{signal::T1}
\end{align}
where $P$ is the transmit power, and $\mathbf N_b$ is the effective noise received at the BS, which includes both the residual self-interference after self-interference cancellation and the additive white Gaussian noise (AWGN).
$\bm \Theta_b, b\in\{1,\ldots, B\}$ is is the BD-RIS scattering matrix at the subframe $b$.
It is worth noting that, unlike cascaded channel estimation in (\ref{signal::uplink3}), where the BD-RIS scattering matrix must be designed for each individual time slot, the proposed channel estimation approach keeps the scattering matrix static throughout each subframe, covering $T$ consecutive time slots. 
This makes the proposed method more practical.
Based on  $\mathbf Y_b$ in (\ref{signal::T1}), the BS then estimates elevation angles at the BS as well as the the elevation and azimuth angles  at the BD-RIS.

\subsubsection{Estimation of elevation angles  $\{\iota_{b,l}\}$ at the BS}
\label{B}
According the structure in (\ref{channel::BS-RIS}),  $\mathbf E_T$ and $\mathbf E_R$ in (\ref{signal::T1}) can be decomposed into three parts as 
\begin{align}
\mathbf{E}_R=\mathbf{B}_R\bm{\Gamma}\mathbf{A}^T,\label{E_R}\\
\mathbf{E}_T=\mathbf{B}_T\bm{\Gamma}\mathbf{A}^T, \label{E_T}
\end{align}
where 
\begin{equation}\label{A_RT}
\small
\left\{\begin{split}
\bm \Gamma & = {\text {diag}} \{ \alpha_{1}, \dots, \alpha_{L} \} \in {\mathbb C}^{L \times L},\\
\mathbf A & = \left[ \mathbf a (\iota_{r, 1},\phi_{r,L}), \dots, \mathbf a (\iota_{r, L},\phi_{r,L}) \right] \in {\mathbb C}^{N \times L},\\
\mathbf B_R & = \left[ \tilde {\mathbf b} (\iota_{b,1}), \dots, \tilde{\mathbf b} (\iota_{ b,L}) \right] \in {\mathbb C}^{M_R \times L},\\
\mathbf B_T & = \left[ \bar {\mathbf b} (\iota_{b,1}), \dots, \bar{\mathbf b} (\iota_{ b,L}) \right] \in {\mathbb C}^{M_T \times L},
\end{split}\right.
\end{equation}
where $\tilde {\mathbf b} = \mathbf b(1:M_R)$ and $\bar{\mathbf b}=\mathbf b(1:M_T)$.

Next, we first specify the procedure to estimate $\iota_{b,l}$ in $\mathbf E_R$.
Substituing (\ref{E_R}) and (\ref{E_T}) into (\ref{signal::T1}), we obtain $\mathbf Y_b = \mathbf B_R\mathbf X_b+\mathbf N_b$, where $\mathbf X_{b}=\bm\Gamma\mathbf A^T\bm \Theta_b\mathbf A\bm \Gamma\mathbf{B}^T_{T}$.
Stacking all $\mathbf Y_b$, $b=\{1,\ldots,B\}$, we obtain $\mathbf Y = [\mathbf Y_1,\ldots,\mathbf Y_B]\in\mathbb{C}^{M_R\times BT}$.

Denote $\mathbf{U}_{M_R}\in\mathbb{C}^{M_R\times M_R}$ as the DFT matrix, and its $(m_1,m_2)$-th element is given as $[\mathbf U_{M_R}]_{m_1,m_2}=\frac{1}{\sqrt{M_R}} e^{-j \frac{2\pi}{M_R} (m_1 - 1) (m_2 - 1)}$.
By left-multiplying $\mathbf Y$ and right-multiplying by   $\mathbf{U}^H_{M_R}$ and $\mathbf S^H$,  we obtain
\begin{equation}\label{Y_DFT}
\small
\bar{\mathbf Y}  = \mathbf U_{M_R}^H \mathbf Y\mathbf S^H =\sqrt{P} \mathbf U_{M_R}^H \mathbf B_R \mathbf X +  \mathbf U_{M_R}^H\mathbf N\mathbf S^H,
\end{equation}
where $\mathbf X=\left[\mathbf X_1,\ldots,\mathbf X_B\right]$ and $\mathbf N=\left[\mathbf N_1,\ldots,\mathbf N_B\right]$.
According to \cite{zhou2022channel}, when $M_R$ approaches infinity, $\mathbf U_{M_R}^H \mathbf B_R \mathbf X$ becomes a sparse vector with $L$ non-zero elements, each corresponding to one of the elevation angles $\iota_{b,l}$.
Let $m_l$ denote the $l$-th non-zero
element of this sparse vector,  the elevation angle can be obtained from $m_{l}$ as \cite{zhou2022channel}:
\begin{equation}\small\label{rough_esti_phi}
\iota_{b,l} =\left\{
\begin{array}{ll}
\arccos \frac{(m_{l} - 1) \lambda}{M_R d}, &~{\text {if}}~ \frac{m_{l} - 1}{M_R} \leq \frac{d}{\lambda}, \\
\arccos \frac{(m_{l} - 1 - M_R) \lambda}{M_R d}, &~{\text {if}}~ \frac{m_{l} - 1}{M_R} > \frac{d}{\lambda}. 
\end{array} \right.
\end{equation}
If $M_R$ is infinite, the estimation of $\iota_{b,l}$ using (\ref{rough_esti_phi}) would be accurate, since the signal components would be perfectly isolated.
However, in practice, $M_R$ is finite, resulting in inevitable estimation inaccuracies. 
Although (\ref{rough_esti_phi}) becomes less accurate when the number of receive antennas is finite, it is important to note that when $M_R$ is large, $\bar{\mathbf Y}$ becomes a sparse matrix, with most of its power concentrated in a few rows and only minor leakage into adjacent rows.
Therefore, a coarse estimation of $\iota_{b,l}$ can still be obtained.
Specifically, by comparing the power of each row in $\bar{\mathbf Y}$, we can identify $\hat{L}$ prominent power peaks, where $\hat{L}$ serves as an estimate of the true number of multipath components $L$. 
Based on the detected $\hat{L}$, we obtain the index of the $l$-th power peak as $\hat{m}_{l}$. By substituting $m_{l}$ with $\hat{m}_{l}$ in (\ref{rough_esti_phi}), a coarse estimation of $\iota_{b,l}$ is obtained.
To get a more accurate estimation, we leverage the angular rotation matrix to reduce the estimation error.
The rotation matrix is denoted as $\bm \Lambda  = {\text {diag}} \left\{ 1, e^{j 2 \pi \nu}, \ldots, e^{j 2 \pi (M_R-1)  \nu} \right\}\in\mathbb{C}^{M_R\times M_R}$, where $\nu\in[ - \frac{1}{2M_R}, \frac{1}{2M_R} ]$ is the rotation coefficient.
By left-multiplying the received signal $\mathbf Y$ with $\mathbf U^H_{M_R}\bm \Lambda^H $ and right-multiplying by $\mathbf S^H$, we obtain
\begin{equation}\label{Y_DFT_rota}
\small
\hat{\mathbf Y}
 = \sqrt{P}\mathbf U_{M_R}^H\bm \Lambda ^H \mathbf B_R \mathbf X  +  \mathbf U_{M_R}^H\bm \Lambda ^H\mathbf N\mathbf S^{H}.
\end{equation}
From (\ref{Y_DFT_rota}), 
the $(m,l)$-th element of $\mathbf U_{M_R}^H\bm \Lambda ^H \mathbf B_R $ is 
\begin{align}\label{Y_DFT_rota_ml}
& [ \mathbf U_{M_R}^H \bm \nu^H \tilde {\mathbf b} (\iota_{b,l}) ]_m\nonumber\\
&= \frac{1}{M_R} \sum_{m_x = 1}^{M_R} e^{- j 2 \pi (m_x - 1) \left(\frac{d}{\lambda} \cos \iota_{b,l} - \frac{m - M_R \nu-1 }{M_R}\right)}.
\end{align}
For each new $\nu$, $\frac{m - M_R \nu - 1}{M_R}$ corresponds to a new angle.
When $\frac{d}{\lambda} \cos \iota_{b,l} - \frac{m - M_R \nu-1 }{M_R}=0$, $[ \mathbf U_{M_R}^H \bm \nu^H \tilde {\mathbf b} (\iota_{b,l}) ]_m$ in (\ref{Y_DFT_rota_ml}) reaches the maximum value.
Hence, for a given ${\hat m}_{l}$, to get the maximum value of (\ref{Y_DFT_rota_ml}), we need to find the best $\nu$.
We then formulate the following optimization:
\begin{equation}\small\label{mu_l}
\hat {\nu}_{l} = \arg \max_{\nu \in \{ - \frac{1}{2M_R}, - \frac{1}{2M_R} + \epsilon, \dots, \frac{1}{2M_R}\}} \parallel [ \hat{\mathbf Y} ]_{{\hat m}_{l},:}\parallel^2,
\end{equation}
where $\epsilon$ is the step length used for searching. 
After determining each $\hat {\nu}_{l}$,  replacing $\nu$ with $\hat{\nu}_{l}$, we obtain a more accurate estimation of the elevation angle at the BS as:
\begin{equation}\label{accurate_esti_phi}
\small
{\hat \iota}_{b, l} = \left\{
\begin{array}{ll}
\arccos \frac{({\hat m}_{l} - M_R \hat{\nu}_{ l} - 1) \lambda}{M_R d}, &~{\text {if}}~ \frac{{\hat m}_l - 1}{M_R} \leq \frac{d}{\lambda},\\
\arccos \frac{({\hat m}_l - M_R\hat{\nu}_{ l} - 1 - M_R) \lambda}{M_Rd}, &~{\text {if}}~ \frac{{\hat m}_l - 1}{M_R} > \frac{d}{\lambda}.
\end{array} \right.
\end{equation}
With $\hat \iota_{b, l}$, the estimation of matrix $\mathbf B_R$ is ${\hat {\mathbf B}_R} = \left[ \tilde{\mathbf b} ({\hat \iota}_{b, 1}), \dots, \tilde{\mathbf b} ({\hat \iota}_{b, \hat L}) \right]$.
Similarly, the estimation of $\mathbf B_T$ is reconstructed as 
 ${\hat {\mathbf B}_T} = \left[ \bar{\mathbf b} ({\hat \iota}_{b, 1}), \dots, \bar{\mathbf b} ({\hat \iota}_{b, \hat L}) \right]$.


\subsubsection{Estimation of the  elevation angles $\{\iota_{r,l}\}$, azimuth angles $\{\phi_{r,l}\}$  at the RIS and channel gain $\{\alpha_l\}$}
\label{A}
Next, we estimate $\left(\iota_{r,l},\phi_{r,l}\right)$ and $\alpha_{l}$. 
Let $\hat{\mathbf B}_R^{\dagger}$  and $\hat{\mathbf B}_T^{\dagger}$ denote  as the pseudo-inverse matrices of $\hat{\mathbf B}_R$ and $\hat{\mathbf B}_T$, respectively.
By left-multiplying the received signal $ \mathbf Y_b$ by ${\hat {\mathbf B}}_R^{\dagger}$ and right-multiplying by $\mathbf S^H{\hat {\mathbf B}}_T^{\dagger}$, we  obtain
\begin{equation}\small\label{approx1}
 \tilde{\mathbf Y}_b=\sqrt{P}\bm \Gamma \mathbf A^T\bm\Theta_b\mathbf A\bm \Gamma  \!+\!   {\hat {\mathbf B}}_R^{\dagger}\mathbf{N}_b\mathbf S^H{\hat {\mathbf B}}^{\dagger}_T.  
\end{equation}
Let $\mathbf V_b=  {\hat {\mathbf B}}_R^{\dagger}\mathbf{N}\mathbf S^H{\hat {\mathbf B}}_T^{\dagger}$ and  $\bm \theta_b\in \mathbb{C}^{N^2\times 1}$ denote the  vectorization of $\bm \theta_b^T$.
According to\cite{zhou2024individual}, it can be observed that the for $1\leq l_m,l_n\leq L$, the $(l_m, l_n)$-th entry of the 
$\tilde{\mathbf Y}_{b}$ is 
\begin{equation}
\small
\begin{split}
&\left[\tilde{\mathbf Y}_{b}\right]_{l_m, l_n} \nonumber \\
&\approx \alpha_{l_m}\mathbf a^T(\iota_{r,l_m},\phi_{r,l_m})\bm\Theta_b\mathbf a(\iota_{r,l_n},\phi_{r,l_n})\alpha_{l_n}+\left[\mathbf V\right]_{l_m,l_n},\nonumber\\
&=\bm \theta^T_b\left(\mathbf a(\iota_{r,l_n},\phi_{r,l_n})\otimes \mathbf a(\iota_{r,l_m},\phi_{r,l_m})\right)\alpha_{l_m}\alpha_{l_n}+\left[\mathbf V\right]_{l_m,l_n},\nonumber\\
&=\bm\theta^T_b\tilde{\mathbf a}\left(\iota_{m,n},\phi_{m,n}\right)\alpha_{l_m}\alpha_{l_n}+\left[\mathbf V\right]_{l_m,l_n},
\end{split}
\end{equation}
where $\tilde{\mathbf a}\left(\iota_{m,n},\phi_{m,n}\right)=\mathbf a(\iota_{r,l_n},\phi_{r,l_n})\otimes \mathbf a(\iota_{r,l_m},\phi_{r,l_m})$.
Stacking all $\bm \theta_b$, $b=\{1,\ldots,B\}$, we obtain $\bm \Phi = [\bm \theta_1^T;\ldots;\theta_B^T]\in\mathbb{C}^{B\times N^2}$.
By collecting $\left[\tilde{\mathbf Y}_{b}\right]_{l_m, l_n}$ in $B$ subframes, we obtain 
\begin{equation}
    \begin{split}
\mathbf{q}_{l_{m}, l_{n}} & =\left[\left[\mathbf Y_1\right]_{l_m,l_n},\ldots,\left[\mathbf Y_B\right]_{l_m,l_n}\right]^T \\
& \approx \bm{\Phi }\tilde{\mathbf a}\left(\iota_{m,n},\phi_{m,n}\right) \alpha_{l_{m}} \alpha_{l_{n}}+\mathbf{n}_{l_{m}, l_{n}},
\end{split}
\end{equation}
where $\mathbf{n}_{l_{m}, l_{n}}=\left[\left[\mathbf V_1\right]_{l_m,l_n},\ldots,\left[\mathbf V_B\right]_{l_m,l_n}\right]^T$.
Hence, the estimated elevation and azimuth angles at BD-RIS and the gain $\alpha_{l_{m}}\alpha_{l_{n}}$ are respectively given by 
\begin{equation}
\label{eleandazi}
    \begin{split}
\left(\hat{\iota}_{m,n},\hat{\phi}_{m,n}\right) & =\arg\max _{\iota_{x,y},\phi_{x,y}} \frac{\left|\mathbf{q}_{l_{m}, l_{n}}^{\mathrm{H}}\bm \Phi \tilde{\mathbf{a}}\left(\iota_{x,y},\phi_{x,y}\right)\right|}{\left\|\mathbf{q}_{l_{m}, l_{n}}\right\|\left\|\bm \Phi \tilde{\mathbf{a}}\left(\iota_{x,y},\phi_{x,y}\right)\right\|}, \\
\widehat{\alpha_{l_{m}} \alpha_{l_{n}}} & =\left(\bm \Phi \tilde{\mathbf{ a}}\left(\hat{\iota}_{m,n},\hat{\phi}_{m,n}\right)\right)^{\dagger} \mathbf{p}_{l_{m}, l_{n}}.
\end{split}
\end{equation}
Let $l_m=l_n$ and  $l_m,l_n\in\left[1,L\right]$ and searching different $\left(\iota_{x,y},\phi_{x,y}\right)$, we can estimate $\left(\hat{\iota}_{m,m},\hat{\phi}_{m,m}\right)$ and extract $\left(\iota_{m},\phi_{m}\right)$ from it.
Besides, we can also obtain $\hat{\alpha}_{l_{m}}=\sqrt{\widehat{\alpha_{l_{m}} \alpha_{l_{m}}}}$  after attaining $\left(\hat{\iota}_{m,m},\hat{\phi}_{m,m}\right)$.
It is worth noting that  there is a phase ambiguity for each $\hat\alpha_{l_m}$, i.e., $\hat{\alpha}_{l_m}$ may be equivalent to $\hat{\alpha}_{l_m} = \alpha_{l_m} $ or $\hat{\alpha}_{l_m} = -\alpha_{l_m}$.
To overcome this ambiguity, we collect more estimations of $\widehat{\alpha_{l_{m}} \alpha_{l_{n}}}, 1\leq l_m,l_n\leq L$ from $L^2$ samples and we could obtain
\begin{equation}
\small
    \underbrace{\left[\begin{array}{cccc}
\widehat{\alpha_{1} \alpha_{1}} & \widehat{\alpha_{1} \alpha_{2}} & \cdots & \widehat{\alpha_{1} \alpha_{L}} \\
\widehat{\alpha_{2} \alpha_{1}} & \widehat{\alpha_{2} \alpha_{2}} & \cdots & \widehat{\alpha_{2} \alpha_{L}} \\
\vdots & \vdots & \ddots & \vdots \\
\widehat{\alpha_{L} \alpha_{1}} & \widehat{\alpha_{L} \alpha_{2}} & \cdots & \widehat{\alpha_{L} \alpha_{L} }
\end{array}\right]}_{\mathbf{D}}=\boldsymbol{\alpha} \boldsymbol{\alpha}^{\mathrm{T}}
\end{equation}
where $\bm \alpha=\left[\alpha_1,\ldots,\alpha_{L}\right]^T$.
To estimate $\bm \alpha$, we exploit the singular value decomposition (SVD)\cite{zhou2024individual}.
Specifically, we first construct a symmetric matrix $\bar{\mathbf D} = \frac{1}{2}\left(\mathbf D+\mathbf D^H\right)$.
Leveraging SVD, $\bar{\mathbf D}$ is decomposed into $\bar{\mathbf D} =\mathbf U_D\bm\Sigma\mathbf U^H_D$.
Let $\sigma_{D,i}$ and $\mathbf u_{D,i}$ denote the $i$-th maximum singular value and the corresponding singular vector, respectively.
Next, we estimate $\hat{\bm \alpha}= \sqrt{\sigma_{D,1}}\mathbf u_{D,1}$.
Again, there exists a phase ambiguity for $\hat{\bm \alpha}$, i.e., $\hat{\bm \alpha}=\bm \alpha$ or $\hat{\bm \alpha}=-\bm \alpha$.
Hence, with given $\hat{\bm \alpha}$, we obtain either $\hat{\bm \Gamma}=\bm \Gamma$ or $\hat{\bm \Gamma}=-\bm \Gamma$.
Finally, after estimating all  parameters, we construct the estimated BS-RIS channel $\hat {\mathbf E}=\hat{\mathbf B}\hat{\bm \Gamma}\hat{\mathbf A}^T$, where $\hat{\mathbf B}=\left[\mathbf b(\hat\iota_{b,1}),\ldots,\mathbf b(\hat\iota_{b,\hat{L}})\right]$, $\hat{\mathbf A}=\left[\mathbf a\left(\hat \iota_{r,1},\hat \phi_{r,1}\right),\ldots,\mathbf a\left(\hat \iota_{r,\hat L},\hat \phi_{r,\hat L}\right)\right]$.
Though the channel $\mathbf E$ has the phase ambiguity, it does not influence the estimation of the cascaded channel.
Assume $\hat{\mathbf E} = -\mathbf E$, then the estimation of RIS-user channel is $\hat{\mathbf h}_k = -\mathbf h_k$ since $\mathbf E$ and $\mathbf h_k$ are coupled in the cascaded channel.
The final estimated cascaded channel is $\hat{\mathbf H}_k=\hat{\mathbf h}_k^T\otimes\hat{\mathbf E}=-\mathbf{h}^T_k\otimes-\mathbf E=\mathbf H_k$.
The BS-RIS channel estimation algorithm
is summarized in Algorithm 1.
\begin{algorithm}
\textbf{Input}:$\mathbf Y_{b}$, $\bm \theta_{b}$, $\hat{\mathbf E}=\mathbf 0_{M\times N}$\; 
 \textbf{Stage 1}: Obtain $\hat L$   and $\hat\iota_{b,l}$ based on (\ref{accurate_esti_phi}), then construct  $\hat{\mathbf{B}}$ based on (\ref{A_RT}). \\
 \textbf{Stage 2}: Obtain the estimated $\left(\hat\iota_{r,l},\hat\phi_{r,l}\right)$ and channel gain $\hat\alpha_{l}$ through the method in  section \ref{A}, then construct $\hat{\mathbf{A}}$ and $\hat{\bm \Gamma}$ based on (\ref{A_RT}).\\ 
\textbf{Output}: Estimated BS-RIS channel  $\hat{\mathbf E}=\hat{\mathbf B}\hat{\bm{\Gamma}}\hat{\mathbf A}^T$. 	
\caption{BS-RIS channel estimation algorithm}
\label{BS-RIS algorithm}
\end{algorithm}


\subsection{RIS-user Channel Estimation}
Given  $\hat{\mathbf E}$, we then estimate the time-varing RIS-user channels $\mathbf h_k$, $\forall k$.
Specifically, user-$k$ transmits pilot signal to the BD-RIS, which is then reflected to the BS so as to estimate each $\mathbf h_k$.
Such procedure requires in total $C$ time subframes and $T\geq K$ time slots in each time subframe.
In each subframe,  we denote the pilot sequence of user-$k$ as $\mathbf{x}_k\in\mathbb{C}^{T\times1}$.
Orthogonal pilots are used for different users, i.e., $\mathbf{x}_k^H\mathbf{x}_{k'}=0$, $k\neq k^{'}$,  $\mathbf{x}_k^H\mathbf{x}_{k}=1$.
In the $c$-th subframe, the scattering matrix of the BD-RIS is fixed to $\bm \Theta^{(2)}_{c} $, the receiving matrix at the BS is denoted as 
\begin{equation}\small
    \mathbf Y^{(2)}_{c}=\sum^{K}_{k=1}\sqrt{P}\mathbf{E}\bm \Theta^{(2)}_{c}\mathbf h_k\mathbf{x}_k^T+\mathbf N^{(2)}_{c},\label{RIS-UE:re}
\end{equation}
$\mathbf Y^{(2)}_{c}\in\mathbb{C}^{M\times K}$  is the receiving matrix and $\mathbf N^{(2)}_{c}\in\mathbb{C}^{M\times K}$ is the noise.
To estimate the channel of user-$k$, we right-multiply $\mathbf Y^{(2)}_{c}$ by
$\mathbf{x}^*_k$, we  obtain
\vspace{-2mm}
\begin{equation}
\small
 \begin{split}
    \mathbf  y_{c,k}=\mathbf Y^{(2)}_{c}\mathbf{x}^*_k&=\sum^{K}_{k=1}\sqrt{P}\mathbf{E}\bm\Theta^{(2)}_{c}\mathbf h_k\mathbf{x}_k^T\mathbf{x}^*_k+\mathbf N^{(2)}_{c}\mathbf{x}^*_k \nonumber\\
    &=\sqrt{P}\mathbf F_c\mathbf h_k + \mathbf w_{c,k},
\end{split}   
\end{equation}
where $\mathbf F_c=\mathbf{E}\bm\Theta^{(2)}_{c}$ and  $\mathbf w_{k,c}=\mathbf N^{(2)}_{c}\mathbf{x}^*_k$.
Since the received pilots at the BS need to be no smaller than the size of $\mathbf h_k\in\mathbb{C}^{N\times 1}$.
In one subframe, 
 $\mathbf y_{c,k}\in\mathbb{C}^{M\times 1}$, we obtain that the number of subframes for the RIS-user channel estimation should satisfy $CM\geq N$, namely, $C\geq\min C =\left \lceil \frac{N}{M} \right \rceil$.
 After $C=\left \lceil \frac{N}{M} \right \rceil$ subframes, we  obtain
\begin{align}
   \tilde{\mathbf{y}}_{C,k} = \sqrt{P}\mathbf F\mathbf h_k+\tilde{\mathbf{w}}_{C,k},
\end{align}
where $\tilde{\mathbf{y}}_{C,k}=\left[\mathbf y_{1,k};\ldots;\mathbf y_{C,k}\right]$, $\mathbf{F}=\left[\mathbf F_{1};\ldots;\mathbf F_{C}\right]$ and $\tilde{\mathbf{w}}_{C,k}=\left[\mathbf w_{1,k};\ldots;\mathbf w_{C,k}\right]$.
With $\hat{\mathbf E}$ obtained in the previous step, we then estimate $\hat{\mathbf F}_c=\hat{\mathbf E}\bm\Theta^{(2)}_{c}$.
Hence, we construct the estimate of $\mathbf F$ with $\hat{\mathbf F}=\left[\hat{\mathbf F}_{1};\ldots;\hat{\mathbf F}_{C}\right]$.
Finally, the LS estimate of the RIS-user channel vector $\mathbf{h}_k$ is 
\begin{align}
    \hat{\mathbf{h}}_k=\frac{1}{\sqrt{P}}\left(\hat{\mathbf F}^H\hat{\mathbf F}\right)^{-1}\hat{\mathbf F}^H\tilde{\mathbf{y}}_{C,k}.
\end{align}
Same approach is applied to all users to obtain $\mathbf h_k$, $\forall k$.

\subsection{Pilot Overhead and Computational Complexity}
\subsubsection{Pilot Overhead}
The proposed individual channel estimation contains two major steps. 
In the first step, the BS-RIS channel is estimated by respectively estimating the elevation angles at the BS and the elevation and azimuth angles at the BD-RIS.
The pilot overhead for estimating the elevation angles at the BS is $L\log(M)$, while the pilot overhead for estimating the azimuth and elevation angles at the BD-RIS is $L\log(N^2)$.
Hence, the minimum pilot overhead for the BS-RIS channel estimation is $L\min\left\{\log(M),~\log(N^2)\right\}$.
In the RIS-user channel estimation, we need $\left \lceil \frac{N}{M} \right \rceil$ subframes, and each subframe needs at least $K$ time slots, in total $K\left \lceil \frac{N}{M} \right \rceil$ time slots.
Hence, for our proposed channel estimation method, in each time frame, we estimate BS-RIS channel once and estimate RIS-user channel $\gamma~(\gamma \geq 2)$ times.
The resulting pilot overhead is $L\min\left\{\log(M),\log(N^2)\right\}+\gamma K\left \lceil \frac{N}{M} \right \rceil$.
\subsubsection{Computational Complexity}
The computational complexity of estimating $\iota_{b,l}$ is $\mathcal{O}(\frac{M^3}{\epsilon})$ while that of $\left(\iota_{r,l},\phi_{r,l}\right)$ and $\alpha_l$ is $\mathcal{O}\left(L^2N^4\right)$. 
Therefore, the overall computational complexity for estimating the BS-RIS channel in the first step is $\mathcal{O}\left(\frac{M^3}{\epsilon}+L^2N^2\right)$. 
In the second step, the computational complexity of using LS is $\mathcal{O}\left(N^3+\left(M+N\right)^2K\right)$. 
Hence, the overall computational complexity is $\mathcal{O}\left(\frac{M^3}{\epsilon}+L^2N^4+N^3+\left(M+N\right)^2K\right)$.

\par 


\section{Numerical Results}
In this section, we evaluate the performance of our proposed individual channel estimation algorithm. Following the simulation setting in\cite{zhou2024individual}, we set $M=80$, $M_R=M_T=\frac{M}{2}=40$, $\gamma = 2$, $K=4$ and the transmit power is $P=20$ dBm.
The number of channel paths between the BS and the BD-RIS is $L=3$ and the number of channel paths between the BS and user-$k$ is $U_{k}=4$.
Channel gains $\alpha_{l}$ and $\alpha_{k,l}$  follow the distribution of  $\alpha_{l}\sim\mathcal{CN}\left(0,10^{PL_B/10}\right)$ and $\alpha_{k,l}\sim\mathcal{CN}\left(0,10^{PL_{k}/10}\right)$ respectively, where  $PL_B$ and $PL_{k}$ denote the large-scale fading coefficients of the BS-RIS and RIS-user channels.
To be specific, the large-scale fading coefficients are denoted as 
$PL=32.4+20\log_{10}(f_c)+10\beta\log_{10}(d)+\xi$ in the dB form, where $f_c$  is the carrier frequency, $\beta$ is the channel path loss exponent, $d$ is the distance of the corresponding channel, and $\xi\sim \mathcal{N}(0,4)$ follows the lognormal distribution\cite{zhou2024individual}.
We set $f_c=28$ GHz,  the distance between BS and BD-RIS is $d_{BR}=10m$, the distance between BD-RIS and user is $d_{RU}=50m$.
Besides, we set $\beta_{l}=2.2$ in $PL_B$ and $\beta_{r,l}=2.2$ in $PL_{k}$.
The noise power is  $\sigma^2=-100$ dBm.
The elevation and azimuth angles are both uniformly distributed in $[0,\pi]$.

We compare our proposed  channel estimation method with the BD-RIS channel estimation algorithm proposed in\cite{li2024channel} based on the LS estimator. 
We investigate the normalized mean square
error (NMSE) performance.
The estimated cascaded channel is $\hat{\mathbf H}_k =\hat{\mathbf h}^T_k\otimes\hat{\mathbf E}$.
The NMSE of $\mathbf H_k$ is 
\begin{equation}
\small
\text{NMSE}=\frac{\mathbb{E}\left\{\sum^K_{k=1}\left \| \mathbf H_k-\hat{\mathbf H}_k  \right \|^2_F \right\}}{\mathbb{E}\left\{\sum^K_{k=1}\left \| \mathbf H_k  \right \|^2_F\right\}}.
\end{equation}
\begin{figure}[t]
\centerline{\includegraphics[width=0.3\textwidth]{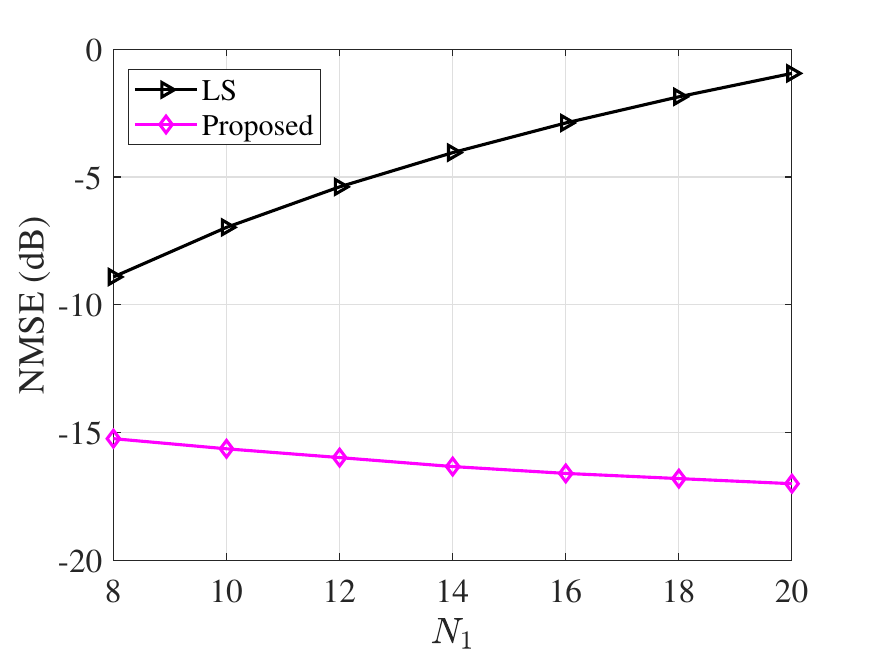}}
\vspace{-3mm}
\caption{The NMSE for the cascaded channel against  with the number of RIS elements with $N_2=N_1$, $p=20$ dBm, $M=80$, $K=4$.}
\label{fig::NMSEvsN}
\vspace{-3mm}
\end{figure}
\begin{figure}[t]
\centerline{\includegraphics[width=0.3\textwidth]{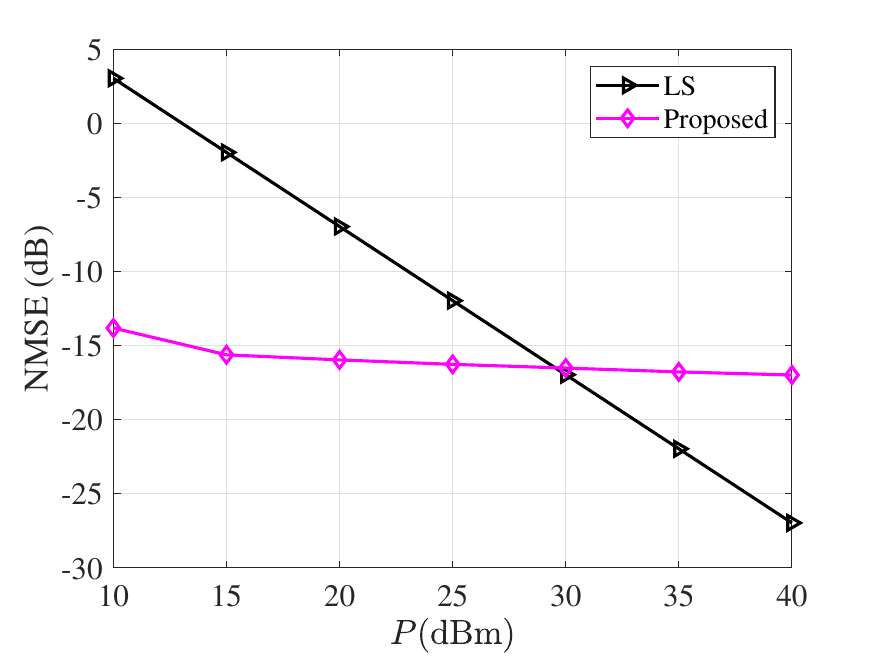}}
\vspace{-3mm}
\caption{The NMSE for the cascaded channel against  transmit power  with $N=100$, $M=80$, $K=4$.}
\label{fig::NMSEvsSNR}
\vspace{-3mm}
\end{figure}
\vspace{-4mm}
\begin{figure}[t]
\centerline{\includegraphics[width=0.3\textwidth]{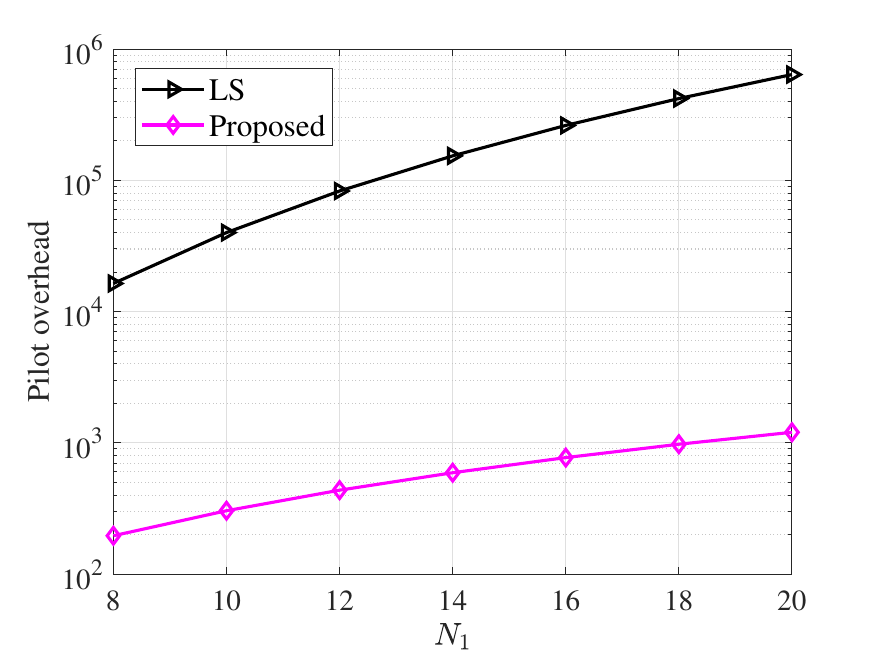}}
\vspace{-4mm}
\caption{The pilot overhead against the nubmer of RIS elements with $p=20$dBm,  $M=80$, $K=4$ and $N_1=N_2$.}
\label{fig::PilotvsN}
\vspace{-3mm}
\end{figure}

Fig. \ref{fig::NMSEvsN} shows the NMSE performance against the number of BD-RIS elements $N=N_1^2$. 
Surprisingly, as the number of BD-RIS elements increases, the NMSE of our proposed method decreases, while the NMSE of the LS-based method increases, highlighting the significant advantages of our approach, especially when the number of RIS elements is large.
The improvement in NMSE for our method is attributed to the increased orthogonality of the steering vectors as the number of RIS elements grows, which enhances the estimation accuracy of $\mathbf E$.
In contrast, the LS-based method does not exploit the structural characteristics of  $\mathbf{E}$, resulting in higher estimation errors as the number of RIS elements increases. 
Compared with the LS-based method, our proposed algorithm consistently achieves better NMSE performance, with an
average decrease of $223$\%.

Fig. \ref{fig::NMSEvsSNR} shows the NMSE performance against transmit power. 
The NMSE of both approaches increases with the transmit power.
The NMSE of the LS-based method is significantly influenced by the transmit power. In contrast, our proposed method is not sensitive to the transmit power.
This is because  our proposed method mainly depends on the structure of the channel, not depending on the transmit power.
Hence,  at low transmit power, our methods can obtain much better NMSE results. However, as the transmit power becomes large, the LS-based method attains much better results.

Fig. \ref{fig::PilotvsN} shows the pilot overhead against the BD-RIS elements. 
It is obvious that our proposed individual channel estimation method significantly reduces the pilot overhead  compared with the LS-based method.
This is because the pilot overhead of the LS method is $KN^2$, proportional to the $N^2$. 
In contrast, the pilot overhead of our proposed method is proportional to  $N$.
Hence, with large BD-RIS elements, our method needs less pilot overhead and attains  more accurate estimation compared with the LS-based method.


\section{Conclusion}
In this paper, we propose a novel individual channel estimation method to separately estimate the BS-RIS channel and RIS-user channel  for BD-RIS.
The key idea is  we estimate the high dimensional  BS-RIS channel over a large time-scale and estimate the RIS-user channel  over a small time-scale.
Moreover, we propose to exploit the sparsity of the BS-RIS channel so as to estimating the elevation angles, azimuth angles, and channel gain separately.
Simulation results show that our proposed   channel estimation framework achieves higher estimation  accuracy compared with the LS-based method especially when the number of BD-RIS elements is large and SNR is medium or low. Most importantly, the pilot overhead is significantly much lower compared with the LS-based method.
\bibliographystyle{IEEEtran}  
\bibliography{reference}

\end{document}